\documentclass[conference]{IEEEtran}
\ifCLASSINFOpdf
\usepackage[pdftex]{graphicx}
\usepackage{array}
\graphicspath{{../pdf/}{../jpeg/}}
\DeclareGraphicsExtensions{.pdf,.jpeg,.png}
\else
\usepackage{algorithmic}
\usepackage{algorithm}
\usepackage[dvips]{graphicx}
\graphicspath{{../eps/}}
\DeclareGraphicsExtensions{.eps}
\fi
\begin{document}
%
\title{Flexible queries in XML native databases}

\author{\IEEEauthorblockN{Olfa Arfaoui$^1$, Minyar Sassi Hidri$^2$}
\IEEEauthorblockA{$^{1,2}$Universit\'e Tunis El Manar\\ Ecole Nationale d'Ing\'enieurs de Tunis\\Laboratoire Signal, Image et Technologies de l'Information\\
BP. 37, Le Belv\`ed\`ere 1002, Tunis, Tunisia\\
$\{$$^1$olfa.arfaoui,$^2$minyar.sassi$\}$@enit.rnu.tn}
}

\maketitle

\begin{abstract}
To date, most of the XML native databases (DB) flexible querying systems are based on exploiting the tree structure of their semi structured data (SSD). However, it becomes important to test the efficiency of Formal Concept Analysis (FCA) formalism for this type of data since it has been proved a great performance in the field of information retrieval (IR). So, the IR in XML databases based on FCA is mainly based on the use of the lattice structure. Each concept of this lattice can be interpreted as a pair (response, query). In this work, we provide a new flexible modeling of XML DB based on fuzzy FCA as a first step towards flexible querying of SSD.
\end{abstract}
\IEEEpeerreviewmaketitle

\section{Introduction}
In order to retrieve data or relevant information (towards user) within XML databases (DB), several querying methods and researches have been proposed \cite{weigel:ECIR}.

They use jointly the content and structure to find relevant information which corresponds to the needs of the end-user. Indeed, the challenge for some of this work is the introduction of flexibility in the elements which satisfy such querying.

Formal Concept Analysis (FCA) consists of inducing, from a binary relation Objects $\times$ Properties, the pairs of subsets {\em {objects}} and {\em {properties}} where the first {\em {objects}} is the maximum of objects satisfying all the properties of the second subset which is the  maximum  subset of properties already satisfied by all objects.

These pairs are called formal concepts where {\em {Object}} (resp. {\em {properties}}) corresponds to the extension (resp. intension) of a given formal concept.

All these formal concepts form a complete lattice. In the original proposal of Wille \cite{wille:ivan}, the relationship is Boolean. That is to say, an object satisfies totally a property or not satisfied it at all.

To remedy this and take into account relationships allowing gradual satisfaction of a property by an object, fuzzy FCA has been introduced \cite{pollandt:springer,belo:mlq}.

In this case, the notion of satisfaction can be expressed by a membership degree $\in [0, 1]$.

Our contribution joins in this context. So, there is an obvious analogy between the binary relation Objects $\times$ Properties characterizing FCA and a binary relation Documents $\times$ Terms characterizing the IR which quickly attracted an orientation for the use of FCA.

In this case, the documents correspond to formal objects and indexing terms to the formal properties \cite{priss:ko}.
Formal concepts thus obtained are interpreted as pairs $\{response\}$, $\{query\}$ where the query corresponds to the intension of a formal concept and the response corresponds to its extension.

The subsumption relation (partial order) between formal concepts can be seen as a relationship of specialization/generalization between queries\cite{messai:inforsid}. It is in this context that is situated our work for the flexible querying of XML documents based on fuzzy FCA for  modeling and querying XML data.

The rest of the paper is organized as follows. Section 2 presents the existing relation between FCA and IR in both  classical (Boolean) and  fuzzy cases. Section 3 presents our pre-querying process for XML native databases modeling. Section 4 gives the general principle of the querying process. Section 5 concludes the paper and highlights future work.
\section{FCA and IR}
\subsection{Basis of Fuzzy FCA}
FCA provides a theoretical framework for learning hierarchies of concepts. This learning occurs from a formal context $K = (O, P, R)$ where $R$ is a binary relation defined completely between a set of objects $O$ and a set
Boolean properties $P(R \subseteq O \times P)$.

A formal context is usually represented by an adjacency matrix. Given an object $x$ and a property $y$, $R (x) = \{y \in P | xRy\}$ is the properties set owned by the object x ($xRy$ means that $x$ has the property $y$) and respectively,  $R (y) = \{x \in O | xRy\}$ is the set of objects with the property $y$.

It was defined in FCA connections between sets $2^{O}$ and $2^{P}$. These connections are called Galois derivation operators.

Galois operator proposed by Wille\cite{wille:ivan} is also called sufficiency operator\cite{deu:tars}. Given $X \subseteq O $ and $Y \subseteq P$, the sufficiency operator, noted $(.)^{\triangle}$, can express all the properties satisfied by all objects in $X$ as:

\begin{equation}
X^{\triangle}=\{y\in P |\forall x \in O (x \in X \Rightarrow xRy)\}
\end{equation}

The dual pair of operators $((.)^{\triangle}, (.)^{\triangle})$ is a Galois connection that can lead to formal concepts.

A formal concept is a pair $(X, Y)$ such that $X^{\triangle} = Y$ and $Y^{\triangle} = X$. The set $X$ (respectively $Y$) is called extension (respectively intension) of the concept.

The set of all formal concepts is naturally equipped with an order relation (denoted by $\preceq$) and defined as: $(X_{1}, Y_{1})\preceq( X_{2}, Y_{2})$ if and only if $X_{1} \subseteq X_{2}$ or $Y_{2} \subseteq Y_{1}$.

This set equipped with the order relation $\preceq$ forms a complete lattice $B (K)$. Meet and Join operators are described by the fundamental theorem of Ganter and Wille \cite{ganter:sv}.

However, in the original proposal of Wille \cite{wille:ivan}, the relationship is considered as Boolean (two-valued) relation. The need to extend the FCA to fuzzy relations have been quickly highlighted to put into account the incomplete relations (imprecise, uncertain, etc.) \cite{djouadi:lfa,messai:ecai}.

In this context, a fuzzy context is described by $K = (L, O, P, R)$ where $R$ is defined as:

\begin{equation}
R: O \times P \rightarrow L=[0,1]
\end{equation}

The generalization of the Galois operator $(.)^{\triangle}$ in fuzzy formal contexts is based on fuzzy implications \cite{belo:mlq}. This generalization is expressed as a function of sets between the subsets of $O (L^{O})$ and $P(L^{P})$. It is given by:

\begin{equation}
X^{\triangle}(y)=\bigwedge_{x\in O}(X(x)\rightarrow R(x,y))
\end{equation}

and

\begin{equation}
Y^{\triangle}(x)=\bigwedge_{y\in P}(Y(y)\rightarrow R(x,y))
\end{equation}

where $\rightarrow$ denotes the fuzzy implication.
\subsection{IR from XML Documents based on FCA}
In IR, the use of FCA lead searchers to consider all objects (resp. properties) as a set of documents (resp. terms).

Several approaches on IR based on the FCA are based on a common point namely the use of the lattice structure of the  formal concepts set \cite{priss:ko,carpineto:jucs,nauer:ecai,dauet:iccs}.

For XML databases, flexible queries are a research topic which has met a growing interest in the querying of SSD. Several techniques have been proposed to address this problem.

We find some relaxation queries \cite{lee:thesis,amer-yahia:edbt}, techniques based on and correlation and techniques based on approximate matching trees \cite{selkow:ipl}.

All these techniques are based on the tree structure of XML document and introduced flexibility according to this last.

Recently some researches are directed towards the use of FCA for modeling an XML document which will facilitate the mining process in such structures.
\section{Pre-querying Process: XML Document Modeling}
\subsection{Using DOM Parser}
We adopt in our approach, the DOM (Document Object Model) parser which uses a hierarchical approach since it loads the XML document in memory as a tree allowing the user easy navigation in the XML document as well as an opportunity to change and to safeguard these updates \cite{olfa}.

More generally through the DOM, our XML document will have the structure of a tree. The nodes of this tree are typed (elements, attributes, text) and are connected by structural relationships such as {\em parent-son} and {\em ancestor-descendant}.

Thus, the leaf nodes are the textual content of the XML document and they have textuel type. The others nodes are internal nodes. Their type is an element one.

For simplifying the rewriting of the nodes  types in the XML tree, the following assumptions are adopted:
\begin{itemize}
\item $NT:$ node whose type is text.
\item $NE:$ node whose type is element.
\end{itemize}
\subsection{Weighting Terms in an XML Document}
The objective here is to evaluate the importance of information in an XML document and to express its relative importance in the granularity of the elements of this document.

For this, the IR-oriented approach is used to extract and index terms according to processes similar to those used in traditional IR. In this cas, the weighting of these terms must be seen in a new angle.

While in traditional IR, the weight of a term seeks to realize, locally, its importance within the document and, globally, in the collection, it is added to structured IR the importance of term at the element which contains it.

In our approach, we assign a fuzzy weight to each text node in the document. The nodes which have a low weight (close to 0) correspond to negligible information (less important than the information provided by other nodes), while a value close to 1 means that the information is considered extremely important including its position in the graph document. Thus, the term weighting is used to measure the importance of a term in an XML document.

This importance is often calculated from considerations and statistical interpretations (or sometimes linguistic). The objective is to find words that best represent the content of a document.

In our case, the nodes to be weighted have the textuel type which is composed of certain number of terms. It is calculated using the following formula:

\begin{equation}
W_{i,n}=tf_{i,n}\times \log(\frac{n_{t}}{nf_{i}})
\end{equation}

where:

\begin{itemize}
\item $W_{i,n}$ is the weight of the term $t_{i}$ in the node $n$ of XML document;
\item $tf_{i,n}$ is the frequency (number of occurrences) of term $t_{i}$ in the node $n$;
\item $n_{t}$  is the number of nodes constituting the XML document;
\item $nf_{i}$ is the number of nodes in the XML document containing the term $t_{i}$.
\end{itemize}

If $n_{j}$ is an element node, the weight $w_{ij}$ is computed by merging weight $w'_{ij}$, ...,  $w'_{ik}$ of $k$ child nodes (text or element) of node $n_{j}$.

While using fuzzy logic, this fusion is interpreted as fuzzy disjunction descriptions of each of these nodes.

For example, if one chooses the maximum function as operator s-norme, we will have $w_{ij} = max (w'_{ij}, ..., w'_{ik})$, which corresponds to the union of the weight associated with child nodes\cite{faessel:egc}.
\subsection{XML Data Processing}
In this section we present some procedures which aim to prepare XML data for querying process. The algorithm \ref{alg1} shows the steps of data extraction from the XML data.

\begin{algorithm}[h!]
\small{
\caption{Data Extraction}
\label{alg1}
\begin{algorithmic}[1]
\REQUIRE XML document
\ENSURE Sets of objects and attributes
\STATE Traverse the XML tree with Ascending traversal
\STATE Extract text data as a set $E_{0}$ (corresponding to level $0$)
\REPEAT
\STATE Extract parent nodes (structural data) as a set $E_{i}$  (corresponding to level $i$)
\UNTIL The root node
\end{algorithmic}
}
\end{algorithm}

Once the sets of leaf nodes and elements are extracted, we can construct the multi-valued formal context where lines correspond to element nodes and columns to the leaf nodes.

The intersection of a row and a column presents the value of the text. In other word, the test value is put in the intersection of the row and the column which present respectively the leaf and the element nodes.

The multi-valued formal context will be transformed to some mono-valued one via conceptual scaling. The algorithm \ref{alg2} shows the steps of fuzzy conceptual classification.

\begin{algorithm}[h!]
\small{
\caption{Fuzzy Conceptual Classification}
\label{alg2}
\begin{algorithmic}[1]
\REQUIRE XML Document, sets of objects (internal nodes) and attributes (leaf nodes) obtained through algorithm \ref{alg1}
\ENSURE Sets of fuzzy concept lattices numbered in the order of the XML tree
\STATE Traverse the XML tree from bottom to top(Ascending traversal)
\FORALL {Parent node $n$ of the XML tree}
\IF {$n$ is a son node (level 1 in the XML tree)}
\STATE Construct the fuzzy context which is composed of the set $E_{0}$ as a set of objects (rows of the matrix) and the set $E_{1}$ as attributes (columns of the matrix) of context
\ENDIF
\IF {$n$ is a parent node (level $i$ in the XML tree)}
\STATE Construct fuzzy context which consists of a set of objects and a set of attributes that correspond to sets $E_{i-1}$) and $E_{i}$.
\ENDIF
\ENDFOR
\FORALL {L-context}
\STATE Affect the coefficient $w(i,j)$ at the intersection of the line $i$ and column $j$
\STATE Compute the coefficient $w_{i,j}$
\FORALL {Obtained context}
\STATE Construct the corresponding fuzzy concept lattice using the algorithm of construction of fuzzy concept lattices Algorithm-INTENT-SET GENERATION \cite{seddoud:umbb}
\STATE Assign a number to each fuzzy concept lattice during its construction in the order they appear in the XML tree
\ENDFOR
\ENDFOR
\end{algorithmic}
}
\end{algorithm}

\begin{figure}[!htbp]
\centering
\includegraphics[scale=0.3]{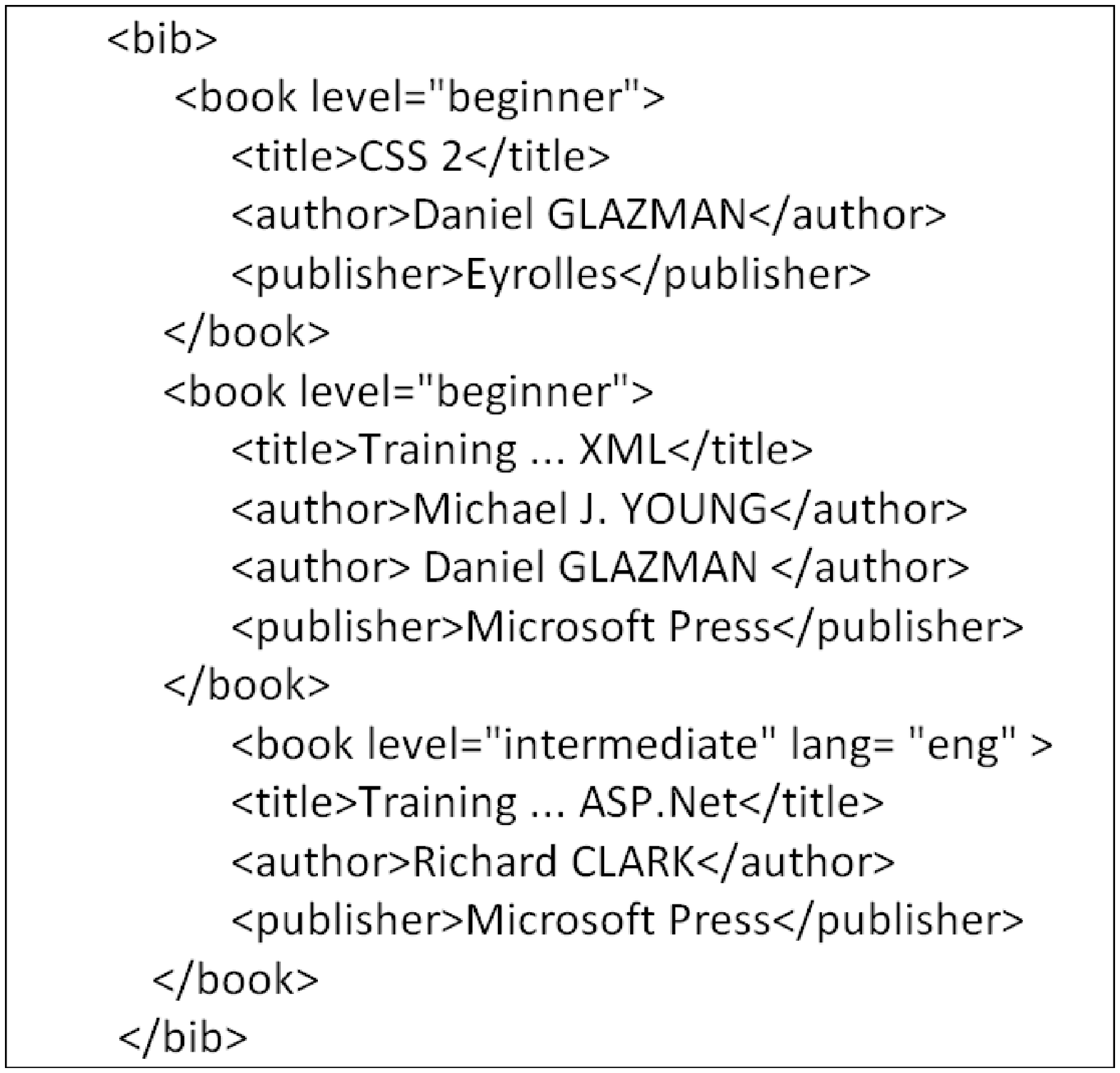}
\caption{An example of XML document}
\label{fig1}
\end{figure}

Figure \ref{fig1} shows an example of an XML document.

The data that we may have in our example are extracted from leaves and nodes as follow: {\em beginner}, {\em CSS 2}, {\em Daniel Glazman}, {\em Eyrolles}, {\em Training ... XML}, {\em Michael J. YOUNG}, {\em Microsoft Press}, {\em Intermediate}, {\em Eng}, {\em Training ... ASP.Net} and {\em Richard Clark}.

In order to simplify the writing we represent these data by the following symbols $\{E1, E2,...,E11\}$.
These data set represent the leaf nodes of the XML tree.

The tree traversal is made with an ascending manner; its aim is to extract the structural data of parent nodes. For example in this example, the first book is a parent node and it has as structural data.

Tables \ref{tab1}, \ref{tab2}, \ref{tab3} and \ref{tab4} present respectively the L-context of the nodes $<book>_{[0]}$, $<book>_{[1]}$, $<book>_{[2]}$ and $<bib>$.

\begin{table}[!htbp]
 \begin{center}
 \caption{L-context of the node $<book>_{[0]}$} \label{tab1}
\footnotesize{
   \tabcolsep = 0.5\tabcolsep
   \begin{tabular}{|c|c|c|c|c|c|c|c|c|c|c|c|}
   \hline
      \multicolumn{12}{|c|}{$<book>_{[0]}$} \\
     \hline
 R&E1&E2&E3&E4&E5&E6&E7&E8&E9&E10&E11       \\
     \hline
  $<level>$&0.17&0&0&0&0&0&0&0&0&0&0      \\
      \hline
   $<lang>$ &0&0&0&0&0&0&0&0&0&0&0     \\
       \hline
     $<title>$&0&0.47&0&0&0&0&0&0&0&0&0      \\
        \hline
      $<author0>$&0&0&0.17&0&0&0&0&0&0&0&0      \\
         \hline
     $<author1>$&1&0&0&0&0&0&0&0&0&0&0     \\
     \hline
     $<publisher>$&0&0&0&0.47&0&0&0&0&0&0&0     \\
     \hline
   \end{tabular}
}
 \end{center}
\end{table}

\begin{table}[!htbp]
 \begin{center}
 \caption{L-context of the node $<book>_{[1]}$} \label{tab2}
 \footnotesize{
   \tabcolsep =0.5\tabcolsep
   \begin{tabular}{|c|c|c|c|c|c|c|c|c|c|c|c|}
   \hline
      \multicolumn{12}{|c|}{$<book>_{[1]}$} \\
     \hline
 R&E1&E2&E3&E4&E5&E6&E7&E8&E9&E10&E11       \\
     \hline
  $<level>$& 0.47 &0 &0 &0 &0&0 &0 &0&0&0&0      \\
      \hline
   $<lang>$ & 0 &0 &0 &0 &0&0 &0 &0&0&0&0     \\
       \hline
     $<title>$& 0 &0&0 &0 &0.47&0 &0 &0&0&0&0      \\
        \hline
      $<author0>$& 0 &0 &0 &0 &0&0.47 &0 &0&0&0&0      \\
         \hline
     $<author1>$& 0 &0 &0.17 &0 &0&0 &0 &0&0&0&0     \\
     \hline
     $<publisher>$& 0 &0 &0 &0&0&0 &0.47 &0&0&0&0     \\
     \hline
   \end{tabular}
}
 \end{center}
\end{table}

\begin{table}
 \begin{center}
 \caption{L-context of the node $<book>_{[2]}$} \label{tab3}
 \footnotesize{
   \tabcolsep = 0.5\tabcolsep
   \begin{tabular}{|c|c|c|c|c|c|c|c|c|c|c|c|}
   \hline
      \multicolumn{12}{|c|}{$<book>_{[2]}$} \\
     \hline
    R&E1&E2&E3&E4&E5&E6&E7&E8&E9&E10&E11      \\
     \hline
  $<level>$& 1&0&0 &0 &0&0 &0 &0.47&0&0&0      \\
      \hline
   $<lang>$ & 1 &0 &0 &0 &0&0 &0 &0&0.47&0&0     \\
       \hline
     $<title>$& 0 &1 &0 &0 &0&0 &0 &0&0&0.47&0      \\
        \hline
      $<author0>$& 0 &0 &1 &0 &0&0 &0 &0&0&0&0.47      \\
         \hline
     $<author1>$& 0 &0 &0 &0 &0&0 &0 &0&0&0&0     \\
     \hline
     $<publisher>$& 0 &0 &0 &1 &0&0 &0.17 &0&0&0&0     \\
     \hline
   \end{tabular}
}
 \end{center}
\end{table}

\begin{table}
 \begin{center}
 \caption{L-context of the root node $<bib>$.} \label{tab4}
      \footnotesize{
      \tabcolsep = 0.5\tabcolsep
          \begin{tabular}{|c|c|c|c|}
   \hline
         \multicolumn{4}{|c|}{$<bib>$} \\
     \hline
 R&$<book>_{[0]}$ &$<book>_{[1]}$ &$<book>_{[2]}$      \\
     \hline
  $<level>$& 0.17 &0.47 &0.47      \\
      \hline
   $<lang>$ & 0 &0 &0.47      \\
       \hline
     $<title>$& 0.47 &0.47 &0.47    \\
        \hline
      $<author0>$& 0.17 &0.47 &0.47       \\
         \hline
     $<author1>$& 0 &0.17 &0       \\
     \hline
     $<publisher>$&0.47 &0.47 &0.17       \\
     \hline
   \end{tabular}
   }
 \end{center}
\end{table}

The complete fuzzy lattice of XML document is constructed based on two or more simple fuzzy lattices. We will consider first the first simple fuzzy lattice. Next, we consider each node of the lattice and we insert other fuzzy lattice.

Nodes of other lattices that carry the attribute of the first lattice are highlighted. Thus, the principle of fuzzy nesting is to combine Simple fuzzy concepts lattices (already obtained), in a single structure called fuzzy nested lattice. Thus, we obtain the L-context of nodes $<book>_{[1]}$ and $<bib>$. It is presented in table \ref{tab5}.

\begin{table*}
 \begin{center}
  \caption{L-context of the nodes $<book>_{[1]}$ and $<bib>$}
  \label{tab5}
 \small{
   \tabcolsep =0.5\tabcolsep
   \begin{tabular}{|c|c|c|c|c|c|c|c|c|c|c|c|c|c|c|c|}
   \hline
      \multicolumn{12}{|c|}{$<book>_{[1]}$} &  \multicolumn{3}{|c|}{$<bib>$}\\
     \hline
 R&E1&E2&E3&E4&E5&E6&E7&E8&E9&E10&E11&$<book>_{[0]}$&$<book>_{[1]}$&$<book>_{[2]}$       \\
     \hline
  $<level>$& 0.92 &0 &0 &0 &0&0 &0 &0&0&0&0&0.17&0.47&0.47      \\
      \hline
   $<lang>$ & 0.47 &0 &0 &0 &0&0 &0 &0&0&0&0&0&0&0.47     \\
       \hline
     $<title>$& 0 &0&0 &0 &0.47&0 &0 &0&0&0&0&0.47&0.47&0.47       \\
        \hline
      $<author0>$& 0 &0 &0 &0 &0&0.47 &0 &0&0&0&0&0.17&0.47&0.47       \\
         \hline
     $<author1>$& 0 &0 &0.17 &0 &0&0 &0 &0&0&0&0&0&0.17&0     \\
     \hline
     $<publisher>$& 0 &0 &0 &0&0&0 &0.47 &0&0&0&0&0.47&0.17&0.47     \\
     \hline
   \end{tabular}
}
 \end{center}
\end{table*}
\section{Towards XML Mining Process}
The general principle of the fuzzy concept lattice querying which  models XML data is introduced above. Once the fuzzy concept lattice is built (using XML data), research in this lattice can follow the following steps:

\textit{\textbf{Step 1:}} Defining of the fuzzy query: this definition is to describe the objects that can form an answer to the question posed by the query, that is to say, one must give the list of attributes that reflect the characteristics of objects to find.

\textit{\textbf{Step 2:}} Inserting the query in fuzzy concept lattices: this is done by incremental construction algorithm which, in our case,  consists of  adding objects whose attributes are provided. In this case, it suffices to consider an object with all the attributes of the query. This is the reason why the request can be presented as a pair (object, attributes) which ensures its inclusion in the fuzzy lattice.

\textit{\textbf{Step 3:}} Pointing the query in the concept lattice: this step is  just after the update of  fuzzy concept lattice after inserting the query concept. As in \cite{messai:inforsid} this research is based on pruning parts of fuzzy lattice while considering the inclusion property between the intentions of formal fuzzy concepts presented by the sub-concept super-concept relationship. The course of this phase is as follows: for a given concept, we examine its intention, if it contains the attributes of the request it is considered and one  examine its super-concepts, if in contrary the intention of this concept does not verify the attributes of the request, it is ignored and so all its super-concepts. Thus, the super-concept to find is that which contains the attributes of the request and also it does not have a super-concept that contains these attributes.

\textit{\textbf{Step 4:}} Researching relevant objects in the vicinity of query concept in the generalized fuzzy lattice. This research must conform to the definition of relevance that is intended to consider only fuzzy concepts containing objects which are relevant to the query. This research is based on two important factors: the first is the relevance criterion to determine the set of objects that meet the search criteria. The second is how to organize the concepts in the lattice to identify relevant concepts.

So, a query describes certainly the ideal object to be sought, the most relevant object to this query is the one that contains all the attributes of the query. This must't neglect objects describing only a part of the application . These objects will be less relevant than the concept verifying all the attributes of the query. First, the relevance of an object to a given query is the number of attributes shared with this query.

Relevant  Objects to a query are those that appear in the query concept they share all the attributes of the query as well as those that appear in the extensions of the concepts of super-concept of the query (provided through the inclusion property between intention concepts in the lattice). So, to ensure research of relevant concepts, one must  locate the query concept, as a first step, and then browse the super- concepts to Top concept: the most general concept lattice. This formalization of the search strategy leads us to define our XML mining algorithm for searching XML data.
\section{Conclusion}
This paper presents a flexible querying approach which deals with XML native DB while basing on based on L-context. It consists in spreading the tree structure of the XML document to a new structure which is the fuzzy concept lattice.

This is a preprocessing step of the XML document that precedes the process of flexible querying of such documents. We tried to decorticate the multi-valued context of XML document in a certain number of mono-valued fuzzy contexts following some rules and steps, this gave rise to generalized fuzzy concept lattices of XML document.

As future work, we intend to conceive and implement our strategy of flexible querying based on fuzzy concept lattice obtained in the first step of this work.

\end{document}